\documentclass[letterpaper, 10 pt, conference]{ieeeconf}  

\usepackage{cite}
\usepackage{amsmath,amssymb,amsfonts}
\usepackage{algorithm}
\usepackage{algorithmic}
\usepackage{graphicx}
\usepackage{subfigure}
\usepackage{textcomp}
\usepackage{xcolor}

\IEEEoverridecommandlockouts                              

\title{\LARGE \bf
Encrypted Data-driven Predictive Cloud Control with Disturbance Observer 
}

\author{Qiwen Li, Runze Gao and Yuanqing Xia$^*$
\thanks{Q. Li, R. Gao and Y. Xia are with School of Automation, Beijing Institute of Technology, Beijing 100081, P. R. China. (Corresponding author: Yuanqing Xia). E-mail address: penguinlee@bit.edu.cn (Q. Li), runze\_gao@bit.edu.cn (R. Gao), xia\_yuanqing@bit.edu.cn (Y. Xia). }%
}

\begin{document}

\maketitle
\thispagestyle{empty}
\pagestyle{empty}

\begin{abstract}
In data-driven predictive cloud control tasks, the privacy of data stored and used in cloud services could be leaked to malicious attackers or curious eavesdroppers. Homomorphic encryption technique could be used to protect data privacy while allowing computation. However, extra errors are introduced by the homomorphic encryption extension to ensure the privacy-preserving properties, and the real number truncation also brings uncertainty. Also, process and measure noise existed in system input and output may bring disturbance. In this work, a data-driven predictive cloud controller is developed based on homomorphic encryption to protect the cloud data privacy. Besides, a disturbance observer is introduced to estimate and compensate the encrypted control signal sequence computed in the cloud. The privacy of data is guaranteed by encryption and experiment results show the effect of our cloud-edge cooperative design. 
\end{abstract}



\section{INTRODUCTION}

Cloud computing provides enormous computing and storage resources for the implementation of control applications, which brings the concept of cloud control systems (CCSs) \cite{xiaFirst, xiaCCS, xiaCurr}. In CCSs, control algorithms are outsourced and executed on cloud platforms to offer control services for local plants. With the development of CCSs, there is an emerging requirement of cloud control for complex systems. However, the complexity and scale of control systems bring new difficulty in designing model-based cloud control laws, since system models are difficult to obtain. As a kind of model-free control approach, data-driven predictive control (DPC) \cite{xiaData} directly computes control sequences based on the input-output data of the system, which avoids the process of system modeling. Therefore, the combination of CCSs and DPC, i.e., data-driven predictive cloud control (DPCC) \cite{gaoDPC, gaoCCS, gaoJSEE}, takes advantage of data storage and computation in the cloud, as well as the model-free manner in control of complex systems, becoming a potential candidate in CCSs. 

However, in DPCC scenarios, the input-output data and control law of systems are stored and computed in the cloud with no data privacy protection, leading to the risk of privacy leakage. To be specific, an eavesdropper could get access to the private system data through communication channel, cloud storage and memory. The eavesdropper could consequently infer the state and model of the system for malicious purposes, such as advanced persistent threat (APT) design and system state tracking. Thus, the privacy issues in DPCC should be seriously considered. 

As a solution, we use homomorphic encryption (HE) approaches to protect data privacy while computing the DPCC control law, since HE schemes allow computations on encrypted data. Specifically, we use CKKS scheme \cite{CKKS}, which is a RLWE-based HE protocol that ensures the privacy of the scheme through introducing errors to satisfy the hardness of the RLWE problem. In CKKS scheme, complex-number vectors are mapped to integer-coefficient polynomials through interpolation, amplification and truncation. Consequently, the addition and multiplication of ciphertext in polynomials are homomorphically equivalent to element-wise addition and multiplication of plaintext in vectors. In this work, we design a privacy-preserving DPCC controller based on CKKS scheme to compute control sequences while keeping system information invisible to potential attackers. 

When performing the privacy-preserving DPCC tasks described above, we should consider the effects on the control quality induced by system noise and uncertainty. Firstly, errors are introduced to the privacy-preserving DPCC procedure through HE scheme. To be specific, errors are introduced to public keys in CKKS scheme to protect the semantic security properties. Moreover, the amplification and truncation procedure bring noises into ciphertexts. Besides, measurement noise, process noise and system uncertainty are ubiquitous in control systems, which consequently influence the control effect of data-driven approaches. 


Hence, disturbance observer (DOB) \cite{LSH, gaoDPC, DOB} is used to guarantee the control accuracy under the uncertainty, including system noise and errors induced by HE scheme. The function of DOB is to estimate the effects performed on a system based on an auxiliary system. 
If estimated, the system uncertainty could be properly compensated with a suitable magnitude. 

Motivated by the above reasons, the main contributions of the privacy-preserving DPCC based on HE scheme are listed as follows:
\begin{itemize}
    \item We design a private DPCC protocol based on CKKS scheme, which preserves the privacy of sensitive system input-output data. 
    \item We apply the DOB technique to estimate and compensate for the uncertainty induced by the HE scheme and system noise under the privacy-preserving DPCC scenario. 
    \item A numerical example shows the effectiveness of privacy-preserving DPCC with DOB, compared to unencrypted non-DOB and encrypted non-DOB conditions. 
\end{itemize}

The remainder of this work is shown as follows. DPCC approaches and their privacy issues are briefly surveyed in Section \ref{work}, based on which we develop a privacy-preserving data-driven control protocol in Section \ref{preliminaries}. In Section \ref{ppDPCC}, a disturbance observer is proposed to compensate for the error induced by encryption and data noise. In Section \ref{example} a numerical example of our proposed method is shown to demonstrate its effectiveness. Section \ref{final} concludes the paper. 

\section{RELATED WORKS} \label{work}

Showing potential in model-free control scenarios, DPC approaches compute the control input directly from the input-output data of the system, and have been widely used in extended situations. \cite{tvDPC} propose a model-free approach for linear parameter-varying systems. A data-driven error model is learned with precollected data in \cite{tjDPC} to achieve accurate position tracking with a robot arm. 

DPC approaches may require extensive data to estimate system models or generate control inputs, in which cases the computation time of system input may become the bottleneck of implementation. Thus cloud computing and distributed computing are gathering more and more attention in DPC tasks for the possibility of computation acceleration by properly utilizing elastic resources in the cloud. \cite{gaoCCS, gaoJSEE} develop a cloud-edge-endpoint DPC prototype, showing the feasibility of cloud-based control systems. To optimize the effort of subspace identification task, which is the basis of data-driven control, \cite{gaoFIS} decomposes the identification algorithm to inter-connected containerized tasks through parallel computing. A further implementation of cloud-edge cooperative DPCC \cite{gaoDPC} uses workflow-based parallel cloud control and edge compensation. 

The privacy of data and models could be leaked through outsourced tasks, since the communication channel and execution environment could be eavesdropped by untrusted third-parties. Therefore, encrypted control approaches have been widely studied since it could simultaneously allow the computation of control signals and preservation of data privacy. Encrypted linear feedback controllers are realized in \cite{HE1}. Moreover, the encrypted realization of more efficient and complex control schemes are proposed to fit integrated cloud scenarios. In \cite{PHEMI}, a privacy-preserving subspace identification approach based on partially HE scheme is proposed. Alexandru et al. \cite{alexEDPC} offer offline and online encrypted cloud control designs, both based on HE, to protect the input-output data of DPC based on a single cloud server. Subsequently a privacy-preserving distributed alternating direction method of multipliers approach is designed to perform the system estimation process in ciphertexts \cite{alexADMM}. 

\section{PRELIMINARIES} \label{preliminaries}

In this section, we sketch the preliminaries of DPC and RLWE-based HE. 

\subsection{Implementation of data-driven predictive control}

We consider a state-space expression of discrete linear time-invariant (LTI) system:
\begin{equation}
\label{statespace}
\begin{split}
	x(k+1) =& Ax(k) + Bu(k) + \epsilon_p, \\
	y(k) =& Cx(k) + \epsilon_s,
\end{split}
\end{equation}
where $x(k) \in \mathbb{R}^n, u(k) \in \mathbb{R}^m, y(k) \in \mathbb{R}^p$ are the state, input and output vector of the system, $\epsilon_p$, $\epsilon_s$ are process noise and measure noise of suitable shapes, respectively. In the following statements, vectors are all viewed as column vectors, except for additional specifications. 

In DPC, we cannot access the specific parameter $A$, $B$ and $C$ mentioned in (\ref{statespace}). Therefore, data-driven approaches are used to infer the system information and perform control task. Specifically, we have the input-output data series of the system through time:
$$\{u(n), y(n), n = 1, 2, ..., T\}. $$

At every time step $k$, we use some slices of the input-output data series as prior information of the system for identification, which are denoted as:
\begin{equation}
\begin{gathered}
u_{f}(k)=\left[\begin{array}{c}
u(k) \\
u(k+1) \\
\vdots \\
u(k+N-1)
\end{array}\right], y_{f}(k)=\left[\begin{array}{c}
y(k) \\
y(k+1) \\
\vdots \\
y(k+N-1)
\end{array}\right], \\
u_{p}(k)=\left[\begin{array}{c}
u(k-N) \\
u(k-N+1) \\
\vdots \\
u(k-1)
\end{array}\right], y_{p}(k)=\left[\begin{array}{c}
y(k-N) \\
y(k-N+1) \\
\vdots \\
y(k-1)
\end{array}\right],
\end{gathered}
\end{equation}
and
\begin{equation}
\label{wp}
v_{p}(k)=\left[\begin{array}{l}
y_{p}(k) \\
u_{p}(k)
\end{array}\right],   
\end{equation}
where the subscript "$p$" and "$f$" indicate "past" and "future", respectively.

Based on the slices shown above, we can fit the implicit system expression with linear regression:
\begin{equation}
\label{regression}
    y_f(k) = L_v v_p(k) + L_u u_f(k) + e(k),
\end{equation}
where $L_v$ and $L_u$ are coefficient matrices to be fit with appropriate shapes that contain system information, $e(k)$ is a noise vector. 

Aiming at sufficiently utilizing prior information, we concatenate the slices of data into the form of Hankel matrix:
\begin{equation}
U_f(k) = [u_f(N) \ u_f(N+1) \ \cdots \ u_f(N+j-1)],
\end{equation}
\begin{equation}
Y_f(k) = [y_f(N) \ y_f(N+1) \ \cdots \ y_f(N+j-1)],
\end{equation}
\begin{equation}
V_p(k) = [v_p(N) \ v_p(N+1) \ \cdots \ v_p(N+j-1)]. 
\end{equation}

Thus the linear regression problem (\ref{regression}) can be viewed as:
\begin{equation}
    Y_f(k) = L_v V_p(k) + L_u U_f(k) + E(k). 
\end{equation}

After solving this linear regression problem, i.e. $L_v$, $L_u$ being obtained, we consider an optimal control problem with the loss function 
\begin{equation}
    J = (r_f(k) - y_f(k))^\top \mathcal Q (r_f(k) - y_f(k)) + u_f(k)^\top \mathcal R u_f(k),
    \label{loss}
\end{equation}
where $\mathcal Q$ and $\mathcal R$ are positive-definite matrices of appropriate shapes, $r_f$ is the reference signal. Problem (\ref{loss}) could be solved by taking derivative of $J$ with respect to $u_f$ after substituting (\ref{regression}) to (\ref{loss}):
\begin{equation}\label{formula}
    u_f(k) = (\mathcal R + L_u^\top\mathcal Q L_u)^{-1}L_u^\top\mathcal Q(r_f - L_v v_p(k)), 
\end{equation}
where $u_f(k)$ is a sequence of predicted control signals. 

\subsection{Lattice-based HE}

HE schemes enable addition and/or multiplication on encrypted data, which is ensured by a homomorphism between ciphertext space and plaintext space \cite{HESurvey}. HE schemes can be divided into three categories \cite{alexEDPC}: partially HE, somewhat HE and fully HE. Partially HE schemes only support addition or multiplication. Levelled or somewhat HE schemes extend the functionality of partially HE and enable both addition and multiplication, with limited times of computation. Fully HE schemes allow infinite times of addition and multiplication, thus support evaluating arbitrary computable functions. Some levelled HE schemes could be converted to fully HE schemes with the use of a refresh algorithm called bootstrapping \cite{boot}.   

In this work, we use CKKS scheme \cite{CKKS, boot}, a typical public key encryption scheme which is levelled homomorphic on complex vectors. CKKS scheme supports addition, finite times element-wise multiplication on real vectors, to protect the privacy of data-driven control. Besides, CKKS scheme utilizes key-switching technique to support advanced operation like element-wise vector rotation and relinearization after multiplication. Also, CKKS scheme supports ciphertext rescaling to control the noise expansion caused by specific operations. 

\begin{figure}[ht]
\includegraphics[width=8.4cm]{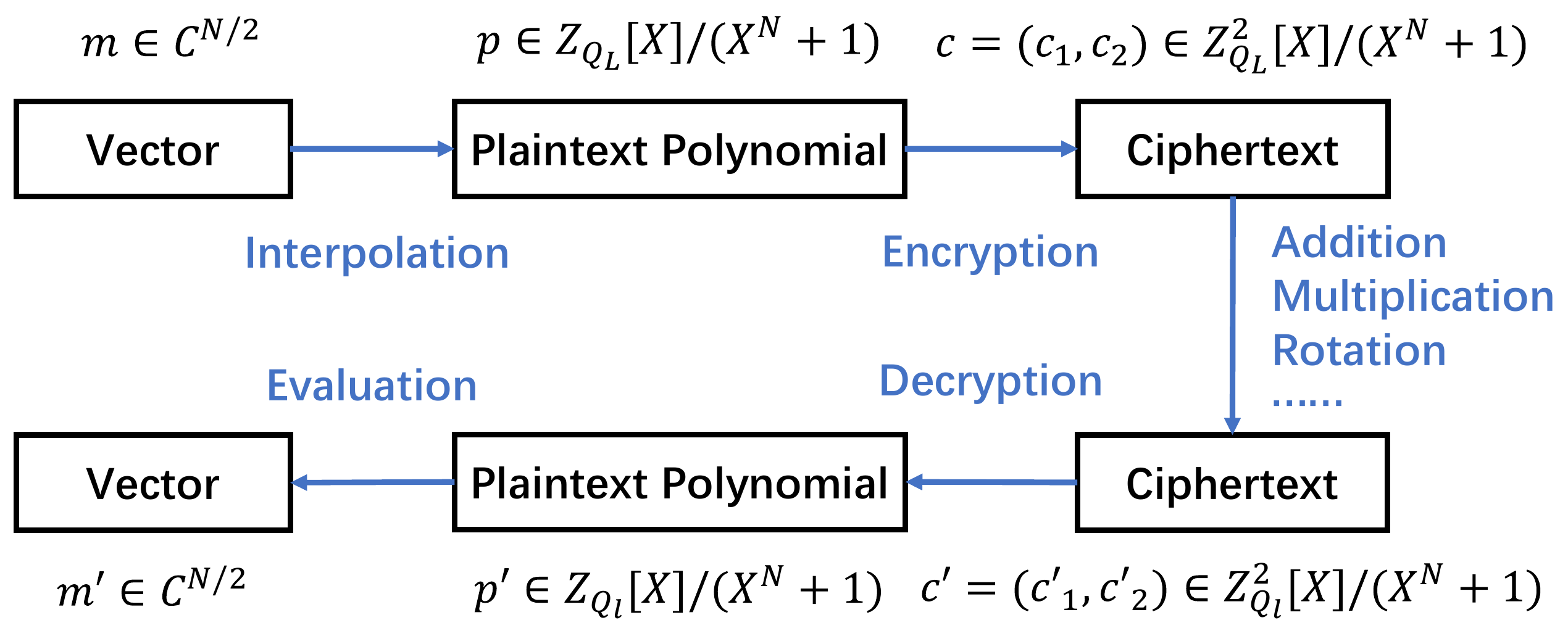}
\caption{A brief description of CKKS scheme.}
\label{ckks}
\end{figure}

A brief description of CKKS scheme is shown in Fig. \ref{ckks}. Denote $N$ be power of 2 and $Q_L$ be a big modulus that equals to the product of a series of positive integers $\{q_0, q_1, ..., q_L\}$. In CKKS scheme, a complex vector $m$ with at most $N/2$ elements is interpolated into a polynomial. Then the embedded polynomial is multiplied by a large scaling factor $\Delta$ and truncated to get plaintext $p$, which is a polynomial in $\mathcal Z_{Q_L}\left[ X \right] / (X^N+1)$, for further encryption. 

The plaintext $p$ will be encrypted into the form of ciphertext $c = (c_0, c_1)$ such that $c_0 + c_1s = p + e\ (mod\ Q_l)$, where $s$ is the secret key and $e$ is the error. Here, ciphertext $c \in \mathcal Z_{Q_l}^2 \left[ X \right] / (X^N+1)$ is denoted to be at level $l$ with $Q_l = \prod_{i=0}^{l} q_i$ for $l = 1,...,L+1$. The plaintext $p$ could be encrypted both by the secret key $s$ and the public key but could be only decrypted with the secret key. The security properties of CKKS scheme are ensured by the hardness of the RLWE problem \cite{HESurvey}. Specifically, all the public keys are in the form of RLWE example $(-as+e, a)$, where random polynomial $a$ and error $e$ safely seal the secret key $s$ according to the hardness of the RLWE problem. Besides, extra public keys in CKKS scheme are available to perform advanced operations like relinearization and rotation to support the design of elaborated computations. 

\begin{figure}[htbp]
\includegraphics[width=8cm]{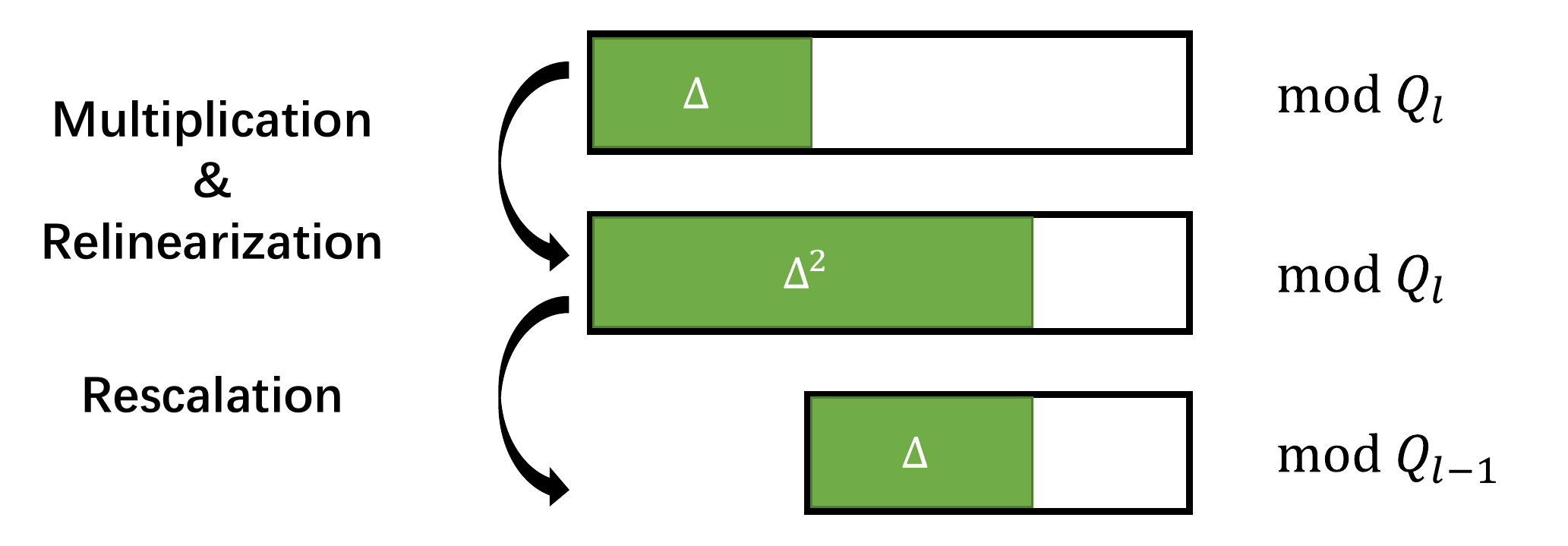}
\caption{Illustrated procedure of the scale limitation in CKKS scheme.}
\label{rescale}
\end{figure}

The noise bound in ciphertexts explodes when performing multiple homomorphic multiplications since the noise is exponentially amplified by the extra scaling factor $\Delta$. As shown in Fig. \ref{rescale}, the multiplication result $c$ at level $l$ could be rescaled by dividing $q_l$, and the level is consequently reduced to $l-1$. Therefore, the noise bound explosion could be reduced to linear expansion, which allows more multiplications to be performed. 

\section{PRIVACY-PRESERVING DPCC DESIGN WITH DOB} \label{ppDPCC}

\begin{figure}[htbp]
\includegraphics[width=8.5cm]{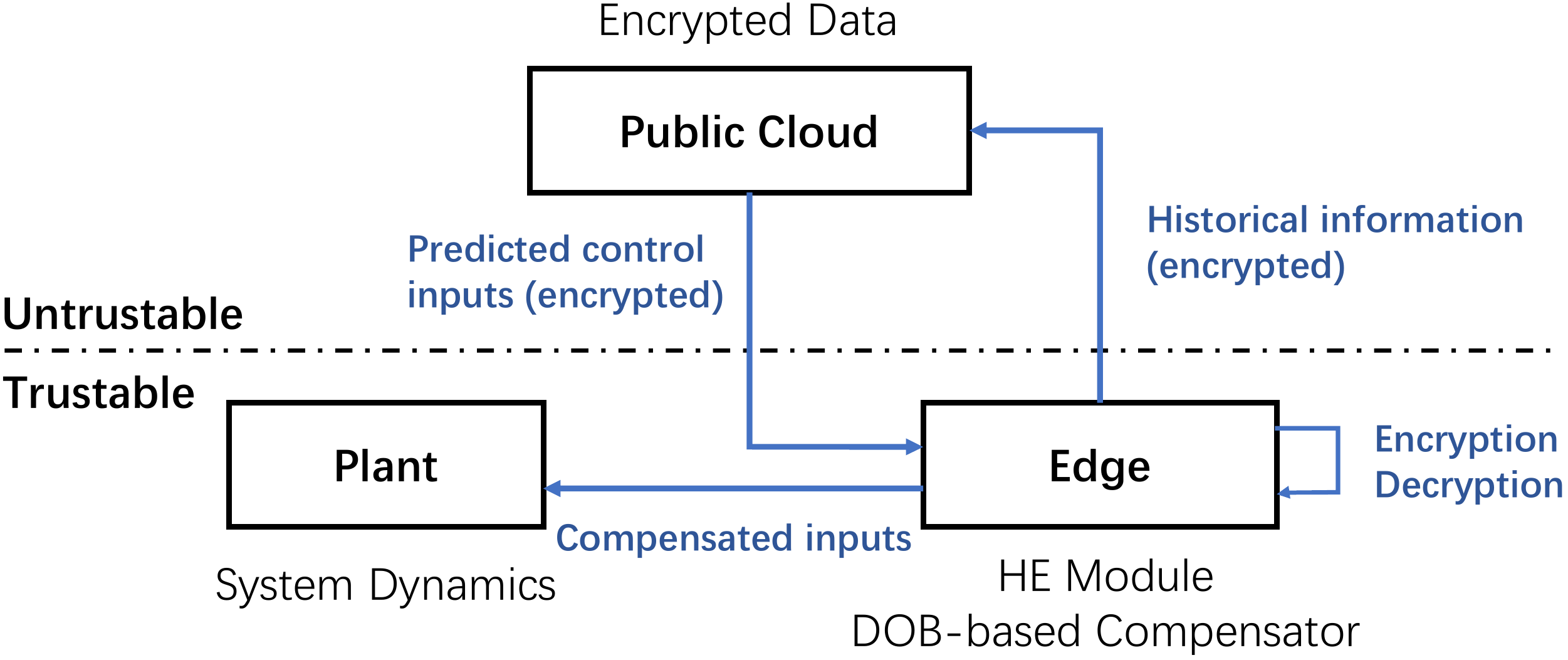}
\caption{Design of privacy-preserving DPCC.}
\label{arch}
\end{figure}

In DPCC scenarios, we assume that the public cloud environment and potential eavesdroppers are honest but curious, which means that they will perform the specified computation or communication correctly, but they want to access the system information to infer the current state and system dynamics. Therefore, the untrusted part placed in the cloud should be encrypted. In this process, the encryption module may introduce new uncertainty. Based on this consideration, the DOB-based privacy-preserving DPCC solution requires the cooperation of three general components: public cloud, trustable edge and plant, respectively. The system design is shown in Fig. \ref{arch}. In the public cloud, an encrypted controller is deployed, maintaining some encrypted matrices to compute encrypted control input sequences. On the trustable edge platform, the HE module is equipped to encrypt and decrypt data, along with a DOB to perform control signal compensation. The plant feeds the modified control input to the system and returns the current output to the edge side. The encrypted data in the cloud controller could be periodically updated to fit the current system dynamics. 

\subsection{Privacy-preserving DPC}\label{AA}

The privacy of the system behavior, including input-output data, should be protected. Similar to \cite{alexEDPC}, an offline privacy-preserving DPC solution is introduced based on CKKS homomorphic encryption scheme.

We could observe that the computation of (\ref{formula}) is realized by specified matrix-vector multiplications. In practice, denote matrix $M_r := (\mathcal R + L_u^\top \mathcal Q L_u)^{-1}L_u^\top \mathcal Q$ and $M_v := (\mathcal R + L_u^\top \mathcal Q L_u)^{-1}L_u^\top \mathcal Q L_v$, which are 2 terms in (\ref{formula}). Since we could compute $L_v$ and $L_u$ in advance, $M_r$ and $M_v$ could be consequently computed offline on a trustable platform, which could be encrypted and uploaded to the cloud, then updated periodically. 

Then, the cloud receives the ciphertexts of $M_r$ and $M_v$, and the control input could be consequently computed:
\begin{equation}
	\label{enccomputation}
	u_f = M_rr_f - M_vv_p,
\end{equation} 
where $v_p$ is the same as in (\ref{wp}) and timestamp $t$ is omitted for convenience. For the efficiency of computation, matrices $M_r$ and $M_v$ would be reused for a given interval and then updated, which is a trade-off in the computation overhead. 

\begin{figure*}
	
	\subfigure[Reformation of matrix and vector. ] {
		\includegraphics[width=8.5cm]{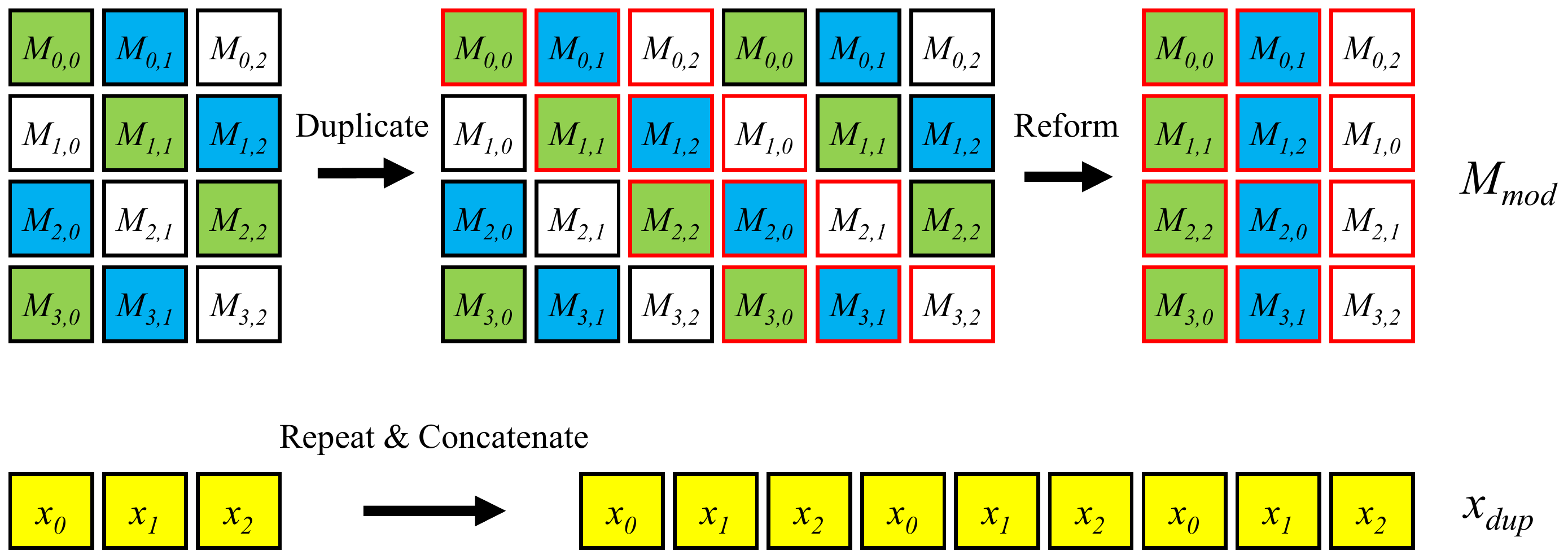}
		\label{pre}
	}
	\subfigure[Matrix-vector multiplication procedure. ] {
		\includegraphics[width=8.6cm]{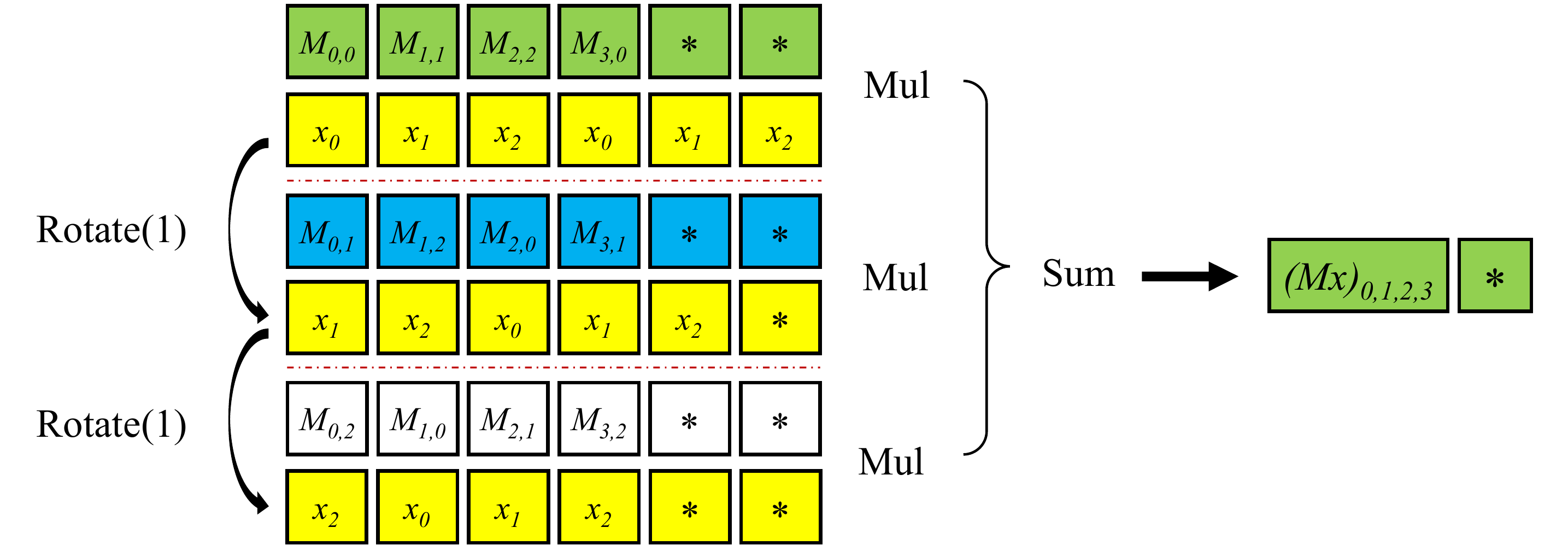}
		\label{mul}
	}

	\caption{Encryption-friendly matrix-vector multiplication: an illustrative example. }
\end{figure*}

Consequently, the computing procedure could be reduced to a matrix-vector multiplication in ciphertext space. Here, a diagonal computation method is utilized to perform the computation \cite{boot}. To implement the encrypted matrix-vector computation $Mx$, the matrix $M \in \mathbb{R}^{K \times L}$ and vector $x\in \mathbb{R}^{L}$ should firstly be rewritten in an encryption-friendly way, which are illustrated in upper part of Fig. \ref{pre}. The modified matrix $M_{mod}$ of matrix $M$ and repeated vector $x_{dup}= \left[ x^\top \ x^\top \ ... \ x^\top \right] ^\top$ of $x$ are provided, which are encrypted and sent to the cloud computing component. 

Denote the encrypted columns of matrix $M_{mod} \in \mathbb{R}^{K \times L}$ as $M_{mod}^{(i)}$, and we need to homomorphically compute matrix-vector multiplication $y = Mx$ in the form of ciphertexts. The matrix-vector multiplication in ciphertext is shown as below:
\begin{equation} \label{encmul}
	y = \sum_{i=0}^{L-1} M_{mod}^{(i)} * rot(x_{dup}, i),
\end{equation}
where the function $rot(x_{dup}, i)$ is the rotation operation supported by the CKKS scheme, meaning that rotating vector $x_{dup}$ $i$ steps to the left. The computation procedures are illustrated in Fig. \ref{mul}. 

Based on above description, the whole encrypted matrix-vector computation procedure is described in Algorithm \ref{algo}. 

\begin{algorithm}[ht]
	\caption{Encryption-friendly matrix-vector multiplication.}
	\label{algo}
	\renewcommand{\algorithmicrequire}{\textbf{Input:}} 
	\renewcommand{\algorithmicensure}{\textbf{Output:}} 
	\begin{algorithmic}[1]
		\REQUIRE
			Matrix $M \in \mathcal{R}^{m \times n}$, vector $x \in \mathcal{R}^{n}$.
		\ENSURE
			Encrypted result of $Mx$.
		\STATE Initialization: build a full zero matrix $M_{mod}$ with the same shape as $M$. 
		\FOR{$i := 0 $ to $ n-1$}
			\FOR{$j := 0 $ to $ m-1$}
				\STATE $M_{mod}[j][i] = M[j][(i + j) \mod n]$. 
			\ENDFOR
		\ENDFOR
		\STATE $x_{dup}$ := Encryption of $\left[x^\top \ x^\top \ ... \ x^\top \right]^\top$. 
		\STATE $M_{mod}^{(0)}, \ ... \ M_{mod}^{(n-1)}$ := Encryption of $M_{mod}$'s columns
		\STATE Compute matrix-vector multiplication through (\ref{encmul}). 
	\end{algorithmic}
\end{algorithm}

 

	

\subsection{DOB and DOB-based cooperative control design}

As analyzed in \ref{preliminaries}, CKKS scheme introduces error to protect its security, meanwhile the amplification and truncation procedures bring error to the system. Besides, the process and measurement noise may also impact the control effect. For reducing the uncertainty and disturbance existed in HE scheme and system dynamics, we adopt the solution in \cite{gaoDPC}, which uses a cloud-edge cooperative control design with a data-driven DOB to estimate the uncertainty and disturbance brought by the cloud. The estimation result obtained by data-driven DOB could be added to the control input for compensation with a proper gain.

Assume that only the first term in the decrypted $u_f$ is fed to the system, which is denoted as $u_c$, as the cloud control signal. We take the nominal input-output relationship into consideration without noise and disturbance:
\begin{equation}
    \label{nominal}
    \begin{split}
        \hat{y}(k+1) = &\quad \sum_{i=1}^{N} \hat{g}_i y(k+i-N) \\
        &+ \sum_{i=1}^{N} \hat{h}_i u(k+i-N) + \hat{b}(k) u_c(k+1),
    \end{split}	
\end{equation}
where $\hat{g}_i$ and $\hat{h}_i$s form the first block row of $\hat{L_v}$ and $\hat{L_u}$, i.e. the disturbed term of $L_v$ and $L_u$, respectively. (\ref{nominal}) is actually the first $p$ rows of the HE implementation of (\ref{regression}). 

If uncertainty and disturbance are considered, the real system dynamics should be:
\begin{equation}
	\begin{split}
		y(k+1) = &\quad \sum_{i=1}^{N} \hat{g}_i y(k+i-N) \\
            &+ \sum_{i=1}^{N} \hat{h}_i u(k+i-N) + \hat{b} u_c(k) + \hat b(k) d(k),
	\end{split}	
\end{equation}
where $d(k) = \Delta u(k)$ is the input disturbance. 

Then, a DOB is introduced with the form
\begin{equation}
	\hat d(k) = P(k) + Ky(k),
\end{equation}
where the disturbance $d(k)$ is estimated by $\hat{d}(k)$, $K$ is the observer amplification matrix to be designed, and $P(k)$ is an auxiliary vector which is updated as below:
\begin{equation}
	\label{update}
	\begin{split}
		P(k+1) = -&K(\sum_{i=1}^N \hat{g}_i(k) y(k+i-N) \\
		+&\sum_{i=1}^N \hat{h}_i(k) u(k+i-N) \\
            +&\hat{b} u_c(k) + \hat b \hat d(k)).
	\end{split}
\end{equation} 

From (\ref{update}), one can obtain 
\begin{equation}
	\hat{d}(k+1) = K \hat b (d(k) - \hat d(k)). 
\end{equation}

Now, define the estimation error as $\Delta d(k) = d(k) - \hat d(k)$ and we have the residue system: 
\begin{equation}
	\Delta d(k+1) = -K\hat{b} \Delta d(k) + d(k+1). 
\end{equation} 

In this system, the edge-compensated input $u_{e}$ is added to the cloud control signal $u_c$, i.e. $u = u_c+u_e$, to get the DPCC cloud-edge co-design. Since the uncertainty caused by HE is viewed as a part of input disturbance, $u_{e}$ is designed to be
\begin{equation}
	u_{e}(k) = -\hat d(k),
\end{equation}
and
\begin{equation}
\begin{split} 
		\hat{d}(k) = \ & K\Big( y(k)-\sum_{i=1}^{N} \hat{g}_{i} (k-1) y(k-N+i-1) \\ 
            &-\sum_{i=0}^{N-1} \hat{h}_{i}(k-1) u(k-N+i-1) \\
            &-\hat{b}(k-1) u_{c}(k-1) \Big)
\end{split}
\end{equation}
when $k = N+1, N+2, ...$. 

When $k = 1, 2, ..., N$, the DOB-based edge compensator do not have enough data in the DPC stage, and $u_e$ could be set to 0 in this time interval, i.e. $u = u_c$. 

\section{NUMERICAL EXAMPLES} \label{example}

We consider a typical 2-order discrete LTI system control problem with parameters
\begin{equation}
	A = \left[\begin{matrix}
		2 & -1 \\
		1 & 0
		\end{matrix}\right],
\end{equation}
\begin{equation}
	B = \left[\begin{matrix}
		1 \\
		0
	\end{matrix} \right],
\end{equation}
and 
\begin{equation}
	C = \left[\begin{matrix}
		0.00014 & 0.00014
	\end{matrix} \right].
\end{equation}

The control input $u$ is clipped between -0.15 and 0.15, and the measure output $y$ is clipped between 0 and 0.4. The system parameters are: $N = 20$, $j = 1000$, $K = 62$, $\lambda = 0.009$. The system state is initialized at $\left[ 0 \ 0\right]^\top$ and the whole control procedure is divided into 2 stages, i.e. data precollection stage and data-driven control stage. In the data precollection stage, the system is controlled through a PID controller with $K_p = K_d = 9$ and $K_i = 3$. The control reference is $y_r = 0.2$ in the first $2N+j=1040$ steps. In the data-driven control stage, $L_v$ and $L_u$ are computed and updated periodically every 50 iterations based on newly collected data. In this stage, the control reference is set to 0.1. 

The whole experiment is realized in a standard Hyper Elastic Cloud Server (HECS) in Huawei Cloud with 2GB RAM and 1 CPU. We implement the private-preserving part of the whole algorithm using the RLWE-based HE library Microsoft SEAL \cite{SEAL}. The security parameter $\lambda$ is chosen to be 128-bit, meaning an encryption scheme could be infiltrated with a probability of $2^{-128}$. The ring dimension is chosen to be 4096, which controls the packing capability of vectors and multiplication depth. The truncation error, which is related to the scaling factor and modulus bits, influences the effect of control. The scaling factor determines the multiplication level, which is bounded by the 128-bit security requirement. The multiplication depth is chosen to be 2, since in this experiment only one multiplication depth is performed in each step. The scaling factor of CKKS scheme is chosen to be $2^{22}$ and $2^{25}$, based on which the influence of floating point number truncation is researched. The process noise and measurement noise are set to be Gaussian with the variance of $0.0027$. 

The experiment is performed to show the control effect of the privacy-preserving DPCC with a DOB-based compensator in three circumstances for comparison, i.e. data-driven control in plaintext, data-driven control in ciphertext with and without DOB-based compensator. 

\begin{figure*}[htbp]
	\centering
	\subfigure[Control results with 22-bit scaling factor.] {
		\includegraphics[width=8.5cm]{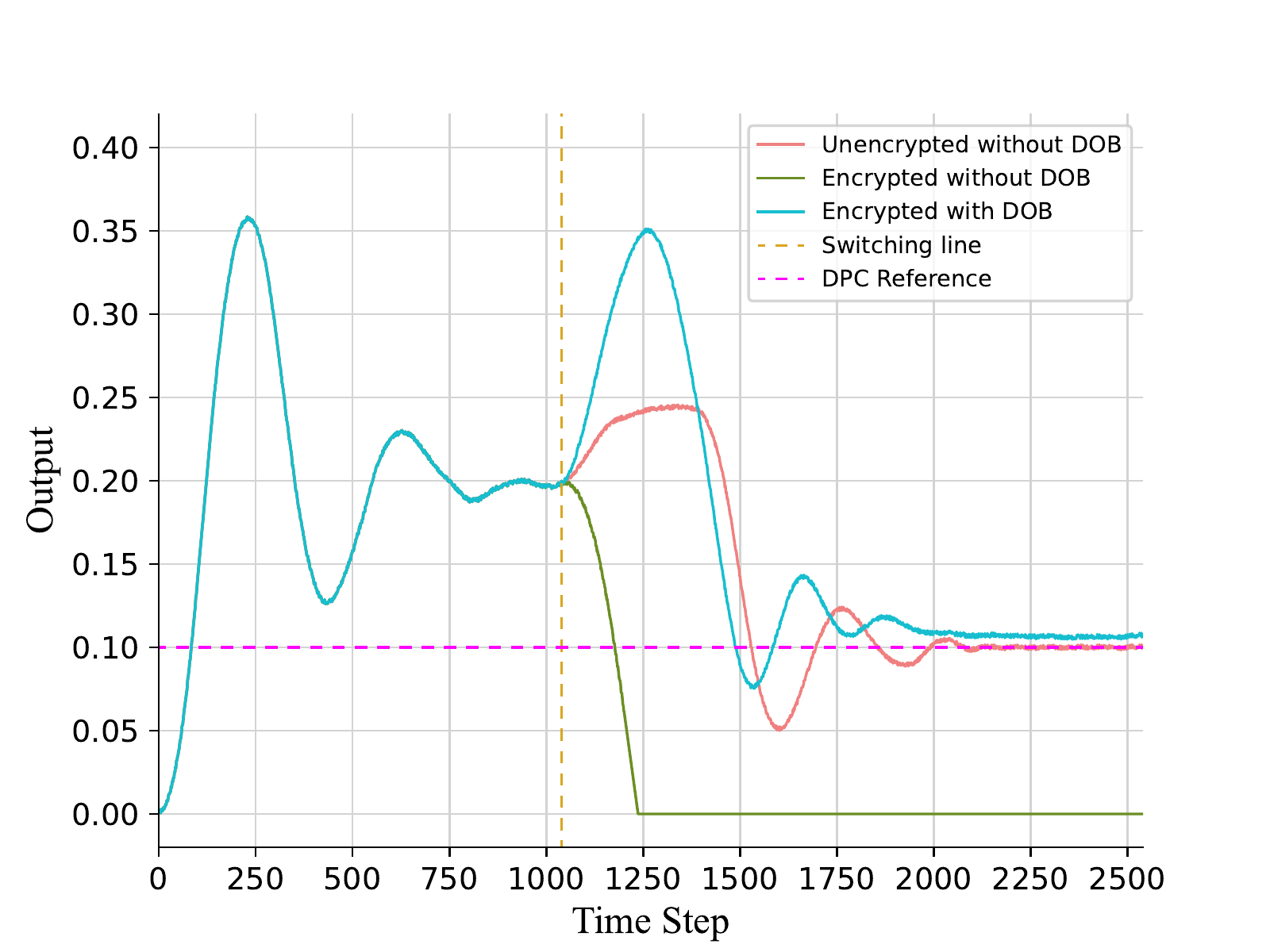}
		\label{fig_22}
	}
	\subfigure[Control results with 25-bit scaling factor.] {
		\includegraphics[width=8.5cm]{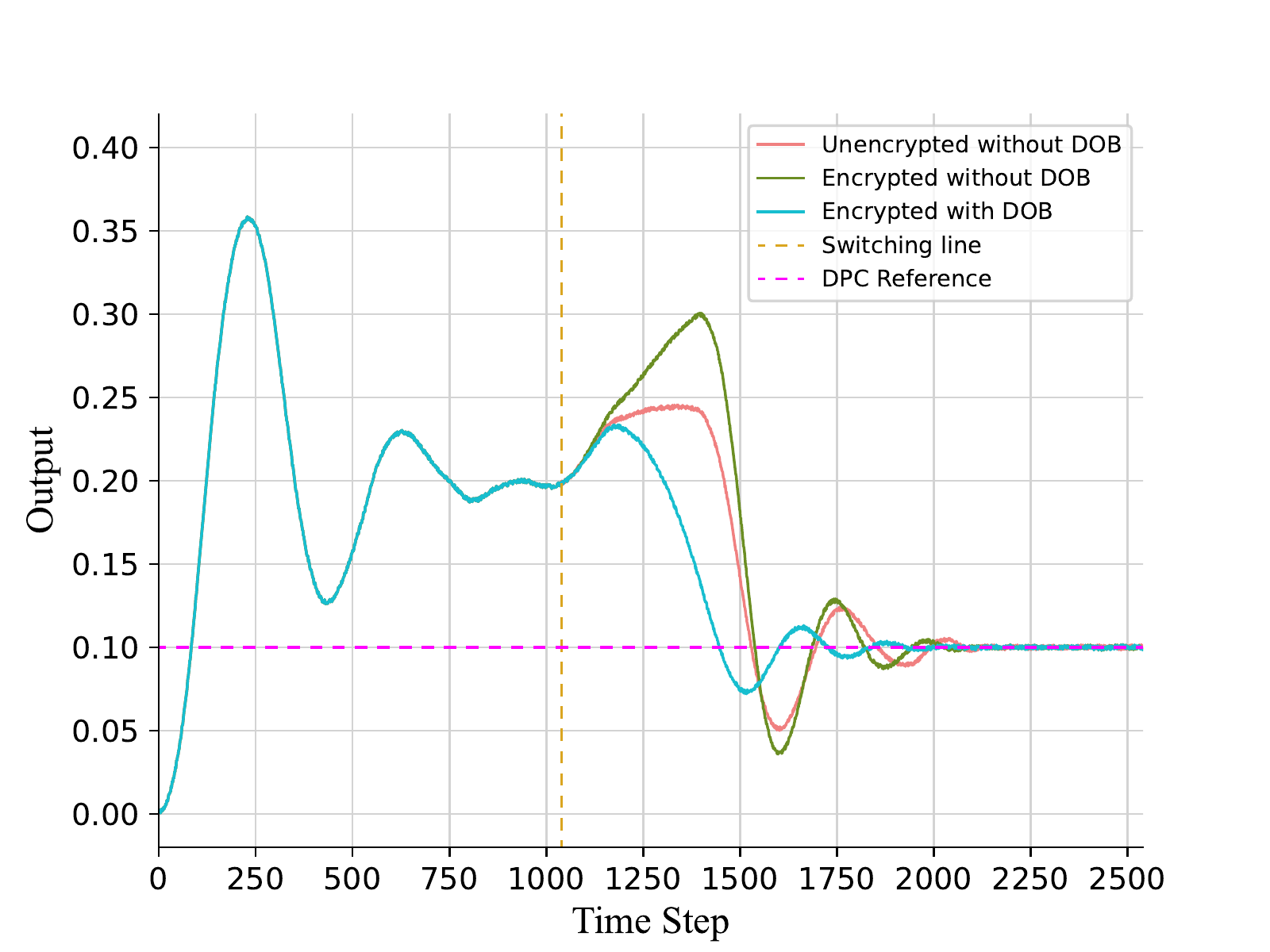}
		\label{fig_25}
	}
\caption{Simulation results of the privacy-preserving DPCC. }
\end{figure*}

The experimental results are illustrated in Fig. \ref{fig_22} and Fig. \ref{fig_25}. As shown in these figures, the DOB-based compensator effectively removes the error induced by system uncertainty, encryption error and external noise. Specifically, in Fig \ref{fig_22}, the scaling factor is set to be $2^{22}$, i.e. about 4 million, which truncates too much information from the plaintext such that compromises the system performance. The system is out of control without compensation. In contrast, DOB-based compensator successfully compensates the uncertainty and disturbance, which improves the control quality. In Fig \ref{fig_25}, the scaling factor is 8 times bigger than $2^{22}$, reducing the truncation error by 8 times, which leads to a similar performance compared to the unencrypted and uncompensated benchmark. In this case, the uncertainty mainly appears in encryption and noise, which could be well estimated and compensated. 

\section{CONCLUSION} \label{final}

In this work, we design a privacy-preserving DPCC solution. Based on HE, we implement a privacy-preserving cloud controller to ensure the data privacy using the CKKS scheme. Also, the uncertainty and disturbance in HE-based control systems are considered, a DOB-based compensator is designed on a trustable edge to estimate and compensate the uncertainty and disturbance. A numerical example shows the effect of our proposed privacy-preserving DPCC design. In the future, the computation efficiency problem of privacy-preserving cloud control solutions would be studied. 

\bibliographystyle{ieeetr}
\bibliography{ref}

\end{document}